\documentclass[]{aa}
\usepackage{graphicx}
\usepackage{times}
\usepackage{latexsym}
\begin{document}
\title{Deprojection of luminosity functions of galaxies in the Coma cluster}
\author{M. Beijersbergen, W.~E. Schaap \and J.~M. van der Hulst}
\institute{Kapteyn Astronomical Institute, P.O. Box 800, 9700 AV
       Groningen,
       The Netherlands}
\mail{beijersb@astro.rug.nl}

\date{Received date; accepted date} 
\abstract{We use a simple analytic model to deproject 2-d
luminosity functions (LF) of galaxies in the Coma cluster measured by Beijersbergen et al.
\cite{beijersbergen_etal}. We demonstrate that the shapes of the LFs
change after deprojection. It is therefore essential to correct LFs
for projection effects. The deprojected LFs of the central area have best-fitting Schechter
parameters of $M^{*}_{\rm U}=-18.31^{+0.08}_{-0.08}$ and
$\alpha_{\rm U}=-1.27^{+0.018}_{-0.018}$, $M^{*}_{\rm B}=-19.79^{+0.14}_{-0.15}$ and
$\alpha_{\rm B}=-1.44^{+0.016}_{-0.016}$ and $M^{*}_{\rm
r}=-21.77^{+0.20}_{-0.28}$ and $\alpha_{\rm
r}=-1.27^{+0.012}_{-0.012}$. The corrections are not significant enough to change the
previously observed trend of increasing faint end slopes with increasing distance to the cluster center. The weighted $U$, $B$, and $r$ band slopes of the deprojected LFs show a slightly weaker steepening with increasing projected cluster
radius. \keywords{Galaxies: clusters:
       individual: Coma (A1656) -- Galaxies: luminosity function --
Galaxies: evolution -- Galaxies: formation.}
} 
\titlerunning{Deprojection of luminosity functions}
\maketitle

\section{Introduction} 

In the past large field studies of the photometric properties of galaxies in the
Coma cluster have been based on photographic plates (e.g. Godwin et
al. \cite{godwin}; Lugger \cite{lugger}). More recent studies
have been made with CCD detectors which provide higher
photometric accuracy. However, a limitation of the use of CCDs has been the
relatively small field of view. Consequently, these studies have been
limited to very small areas, ranging from $\sim50$ arcmin$^2$
(e.g. Bernstein et al. \cite{bernstein_etal}; Mobasher \& Trentham
\cite{mobasher_trentham}) to 1500 arcmin$^2$ (e.g. Lobo et
al. \cite{lobo_etal}), mainly focused on the cluster center. These small fields can not be used to study the
dependence of the luminosity function (LF) on the projected distance from the cluster
center. Furthermore, lacking the photometric data at larger distances from the core
it is not possible to correct a central LF for projection
effects. Accordingly, the published LFs of galaxies
in the Coma cluster are all projected and are,
therefore, luminosity distributions rather than actual LFs. It is important to test how well
these correspond to the actual deprojected LFs.
 
Recently, Valotto et al. (\cite{valotto_etal}) used numerical simulations to analyse systematic effects in the determination of the galaxy luminosity
function in clusters. They find that projection effects conspire to
produce artificially steep faint end slopes. Beijersbergen et al. (\cite{beijersbergen_etal}, hereafter BHDH) presented statistical  $U$, $B$
and $r$ band projected LFs for galaxies in the Coma cluster in five annuli out to a projected radius of 1.4 degrees for the $B$ and $r$
bands and
0.74 degrees for the $U$ band. In total an area of 5.2
deg$^2$ in $B$ and $r$ and 1.3 deg$^2$ in $U$ has been surveyed. This data offers the interesting possibility to deproject actually observed LFs and to obtain true 3-d LFs.

In this \emph{Research Note} we use the information of all annuli to 
deproject the 2-d LFs. Our study extends the work in
BHDH by inferring for the
first time deprojected LFs as a function of projected distance from
the cluster center. In addition, we study how the shapes
of the LFs are affected by projection by measuring the faint end
slopes as a function of distance to the cluster center and compare this with results from numerical simulations. Throughout this \emph{Research Note}
the projected LFs are taken from BHDH and are called 2-d LFs. Deprojected LFs are named 3-d LFs. 

\begin{figure}[!t]
\includegraphics[height=8cm, width=7.35cm]{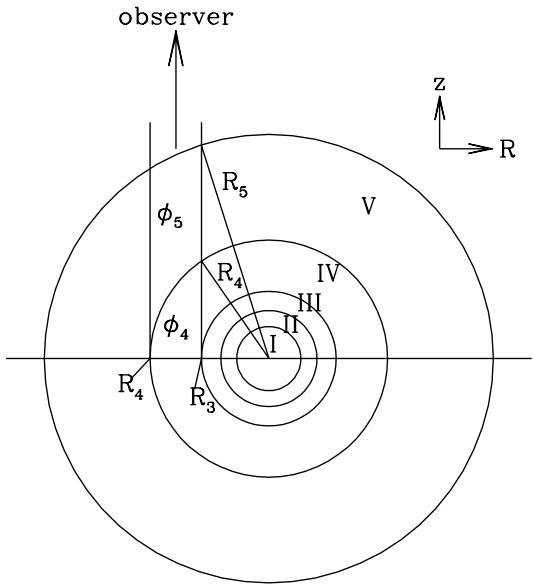}
\caption{Outline of the model used to deproject 2-d luminosity
functions. The figure can be seen as a cut through a sphere and as its projection as seen by the observer. The five circles make up the projected annuli as used in
Beijersbergen et al. \cite{beijersbergen_etal}. Radii are: I: 0-0.2,
II: 0.2-0.3, III: 0.3-0.42, IV: 0.42-0.74 and V: 0.74-1.4 degrees ($1\degr$ = 1.22 $h_{100}^{-1}$ Mpc at Coma
distance).}
\label{3dplot}
\end{figure}

\section{Deprojection of luminosity functions}

Extensive literature exists on the uncovering of three-dimensional
structure from observations of projected quantities (see e.g. Zaroubi
et al. \cite{zaroubi_etal} and references therein). Usually methods
have been described
which were designed for a specific type of observation, in particular
X-ray measurements. Here we use a very simple, analytic geometric
deprojection scheme (Fabian et al. \cite{fabian}). This
scheme has mainly been used for the analysis of X-ray observations, but in
principle it is suitable for any type of observations. We use this projection scheme to deproject the observed 2-d luminosity
distributions of the galaxies in the Coma cluster to real, 3-d, luminosity functions. This scheme consists of assuming that the galaxies
are symmetrically distributed in a sphere. This sphere is
divided up in a number of concentric shells, in which the galaxy density
per magnitude is assumed to be constant. The value of the 3-d
luminosity function in a particular shell can then be found by
subtracting the contributions from the other shells due to projection.
In Fig.~\ref{3dplot} we show an outline of this model. 

The number of projected galaxies per magnitude $m$, N$_{i,m}$, in
annulus $i$, and
the galaxy density per magnitude, $\phi_{m}$, are
related as:

\begin{eqnarray}
\lefteqn{N_{i,m}=\phi_{i,m}~F(R_i,R_{i-1},R_i)~+\sum_{j=i+1}^{5}\phi_{j,m}}
\nonumber\\
&\Bigg\{[F(R_j,R_{i-1},R_i)-F(R_{j-1},R_{i-1},R_i)]\Bigg\}
\label{endeq}
\end{eqnarray}

where R$_i$ are the radii of the annuli and the function F is defined by:

\begin{equation}
F(\alpha,\beta,\gamma)\equiv
\frac{4}{3}~\pi~\alpha^3\left[{(1-\beta^2/\alpha^2)}^{3/2}-{(1-\gamma^2/\alpha^2)}^{3/2}\right]
\end{equation}

\begin{figure*}[!t]
\center
\includegraphics[width=12cm, keepaspectratio]{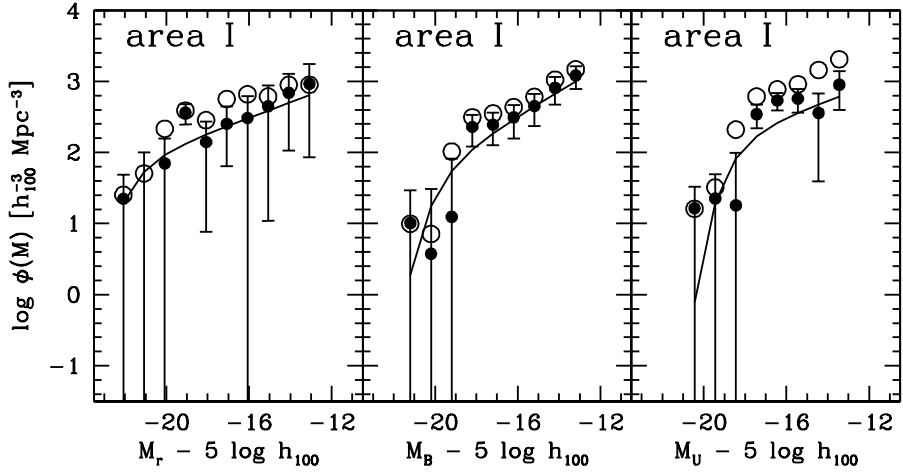}
\caption{Two and three dimensional luminosity functions of area I compared. Open
symbols represent 2-d luminosity functions and solid symbols represent
3-d luminosity functions. The solid lines represent best-fitting
Schechter functions to the 3-d data. For clarity the error bars of the
2-d luminosity functions are not shown.}
\label{corelfscompare}
\end{figure*}

\begin{figure*}[!t]
\center
\includegraphics[width=12cm, keepaspectratio]{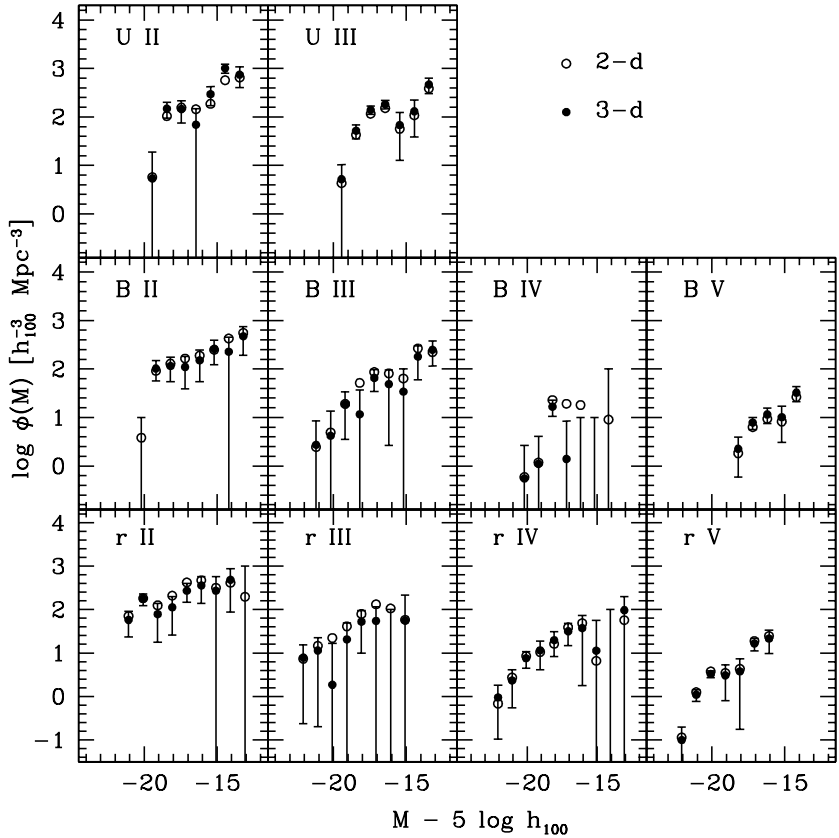}
\caption{Two and three dimensional luminosity functions of areas II to
V compared. Open
symbols represent 2-d luminosity functions and solid symbols represent
3-d luminosity functions. For clarity the error bars of the
2-d luminosity functions are not shown.}
\label{mosaiccompare}
\end{figure*}

With the available 2-d information we can solve for all
$\phi_{i,m}$, i.e. the true 3-d LFs. The $N_{i,m}$ are dependent on the areas of the annuli
which are accurately known. However, the outermost annuli are not
entirely filled by data: annulus V is filled
for $\sim80$\% in $B$ and $r$ and annulus IV is filled for $\sim45$\%
in $U$. We omitted annulus IV for the $U$ band and corrected $N_{5,m}$
for the $B$ and $r$ bands by assuming a constant projected density
throughout this annulus. The function F is solely dependent on geometry and has
values of order $10^{-3}$ to $10^{-1}$. Figs.~\ref{corelfscompare} and
\ref{mosaiccompare} show both the 2-d (open symbols) and
3-d (solid symbols) LFs for all bands and annuli (all magnitudes are
corrected for atmospheric extinction). The error bars of the
2-d LFs are not shown
for clarity, but are somewhat smaller than the 3-d error bars. The 2-d errors are estimates of the uncertainty in the
subtraction of the number of field galaxies. These estimates are based
on very large
control fields and therefore very accurate (see
BHDH for an extensive discussion). The 3-d errors have been calculated
using Eq.~\ref{endeq}. The corresponding errors are
therefore dependent on each other and fully determined by the uncertainty
in the subtraction of the number of field galaxies. The vertical scales represent volume densities and the
2-d LFs are scaled as to match the bright ends of the 3-d LFs. This simplifies the comparison of their shapes. In annulus V the galaxy density is
much lower than in the center, but still higher than in the
field. That is to say, the derived 3-d $B$ and $r$ band LFs still suffer from (minor)
projection effects due to cluster galaxies at radii even larger than
1.7$h^{-1}_{100}$ Mpc. The contamination by projection for the 3-d $U$ band LFs is larger since there we miss photometric information for
radii larger than 0.5$h^{-1}_{100}$ Mpc.

\begin{figure}[!hb]
\includegraphics[height=8cm, width=8cm]{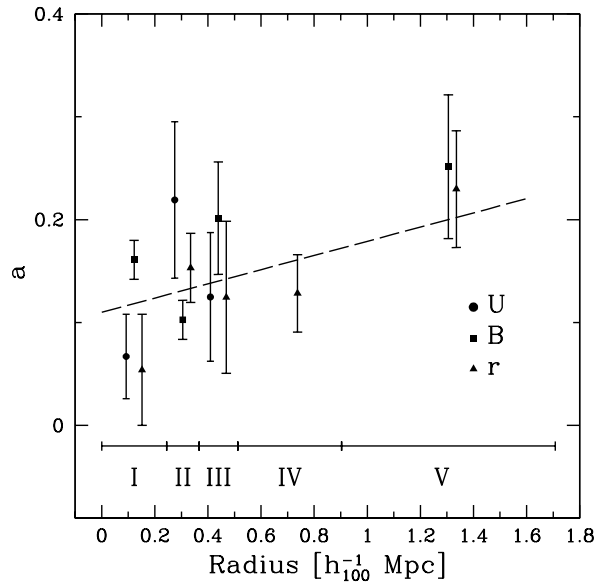}
\caption{Power law slopes $a$ as a function
of cluster radius. The $U$ and $r$ band slopes are offset in radius by
-0.03 and +0.03, respectively. The faint ends of the LFs become steeper
towards the cluster outskirts. The overall trend is indicated by the
dashed line.}
\label{slope_radius}
\end{figure}

\section{Discussion}

In the literature LFs of galaxies in clusters are usually given in units of
number of galaxies per magnitude per area. What actually is measured
is a luminosity distribution, denoted here as 2-d LF. A luminosity \emph{function} of a sample of
galaxies has units of galaxies per magnitude per volume. For
observations of the entire projected volume of a cluster the two have
the same shape since for all luminosities the same volume is
sampled. BHDH studied changes in the LF as a
function of distance to the cluster center. However, they did not
correct for the fact that fore- and background galaxies, residing in the
outer parts of the cluster, are projected onto the central cluster
regions. The question is how much their results have been influenced by this
effect. Below we study how the shapes
of the luminosity distributions given in BHDH are affected by projection effects.

In the outermost annulus the projected area is converted to volume
through a factor depending on the adopted geometry (radii), i.e. for
this annulus the
shapes of the 2-d and 3-d LFs remain
identical. From all LFs of the inner annuli the contributions to the
galaxy counts from surrounding shells is subtracted using the model
described in Sect. 2. The impact on the
resulting LFs can be seen in Figs.~\ref{corelfscompare} and
\ref{mosaiccompare}. When the 2-d and 3-d LFs are compared binwise the
differences, generally, are small. Yet, the error bars become larger
and it should be kept in mind that the scale is logarithmic. Our result can be
explained by the small values of the F integrals in Eq.~\ref{endeq}
combined with the density profile of the cluster. In all expressions
we find combinations of relatively high values of the integrals times low densities, and vice versa, resulting in a minor
net correction term. However, this correction term is different for
each magnitude bin causing the shapes of corrected and uncorrected LFs
to be different. Below, we quantify the differences by studying
the change in the shapes of the 3-d LFs as a function of position in the
cluster. The results are compared to the exact same analysis for the
2-d LFs.

\subsection{Dependence of luminosity functions on radial distance from the cluster center}

BHDH found that the 2-d central LFs could be well
represented by Schechter functions. In Fig.~\ref{corelfscompare} the
best-fitting Schechter (\cite{schechter}) functions of the 3-d LFs are
shown as solid lines. In Table~\ref{alpha} we compare the best-fitting Schechter parameters, $\alpha$ and $M^*$, for the 2-d and 3-d case. For the $r$ and $B$ bands the
Schechter fits are again reasonable representations of the data,
though the best-fitting parameters have changed. The $U$ band fit remains a poor representation of the data.  

\begin{table}
\caption{Comparison of best-fitting Schechter parameters for 2-d and
3-d LFs. $M^*$ is not corrected for extinction.}
\setlength{\tabcolsep}{1.2mm}
\begin{tabular}{lcccc}
Filter&\multicolumn{2}{c}{2-d}&\multicolumn{2}{c}{3-d}\\
&$\alpha$&$M^*$&$\alpha$&$M^*$\\
\hline
\\
U&$-1.32^{+0.018}_{-0.028}$&$-18.60^{+0.13}_{-0.18}$&$-1.27^{+0.018}_{-0.018}$&$-18.31^{+0.08}_{-0.08}$\\
\\
B&$-1.37^{+0.024}_{-0.016}$&$-19.79^{+0.18}_{-0.17}$&$-1.44^{+0.016}_{-0.016}$&$-19.79^{+0.14}_{-0.15}$\\
\\
r&$-1.16^{+0.012}_{-0.019}$&$-20.87^{+0.12}_{-0.17}$&$-1.27^{+0.012}_{-0.012}$&$-21.77^{+0.20}_{-0.28}$\\
\end{tabular}
\label{alpha}
\end{table}

One of the main results of BHDH was the finding of a significant steepening of
the faint end slopes with increasing cluster radius. A straight line
fit ($sr+b$), where $r$ is the clustercentric radius, to the weighted $U$, $B$ and $r$ band slopes at each radius gives $s=0.11\pm0.02$ Mpc$^{-1}$ for the 2-d LFs. In order to compare this with the situation
where all 2-d LFs have been corrected for projection we fitted power laws ($b~10^{aM}$) to the
faint ends of the 3-d LFs in Figs.~\ref{corelfscompare} and
\ref{mosaiccompare}. The infalling group around NGC 4839 (Neumann et
al. \cite{neumann_etal}) does not have a significant influence on any of the derived LFs. The LFs were fitted for $M_{\rm r}>-20$, $M_{\rm B}>-20$ and
$M_{\rm U}>-19$. We omitted the slope of the $B$ band LF for area IV
since there the LF is not well constrained. The results are shown in
Fig.~\ref{slope_radius}. A straight line fit to the weighted slopes of
the 3-d LFs at each radius
gives $s= 0.07\pm0.04$ Mpc$^{-1}$ and is indicated by the dashed
line. Although this is a somewhat weaker trend than derived from the
2-d results, it is different at $1\sigma$ only and the trend is still
clearly there. 

Valotto et al. (\cite{valotto_etal}) used a deep mock catalogue to analyse
systematic effects in the determination of the galaxy luminosity
function in clusters. They find that projection effects produce artificially steep faint end slopes. The cluster core suffers most from objects that are projected onto it. The LFs of area I shows similar ($r$ and $B$ band) or shallower ($U$ band) faint end slopes when they are corrected for projection. We conclude that the impact of
projection effects on the shapes of the 2-d LFs in Coma is not as severe as
their simulations suggest. However, the 2-d LFs derived for relatively small
areas suffer from projection effects and should therefore be properly corrected before one can reliably
interpret their shapes.

\section{Summary and Conclusions}

In this \emph{Research Note} we have used a simple analytic model
to deproject 2-d LFs of cluster galaxies. We have applied this to the
LFs of the galaxies in the Coma cluster measured by BHDH. Under the
assumptions of a spherically symmetric galaxy distribution and a
constant galaxy density in each shell the 2-d LFs have been corrected for
projection effects. The main results are summarized by Figs.~\ref{corelfscompare} and
\ref{mosaiccompare}. Projection effects affect the shapes of the 2-d LFs
and in general make the faint end slopes artificially steeper, although the differences
are small. The alterations are not significant enough as to arrive at different conclusions
compared to those inferred from 2-d LFs. Furthermore, the previously derived
steepening of the faint end slopes with increasing cluster radius is
only slightly weakened when projection effects are taken into account. Overall, we have demonstrated by using a
simple model that before interpreting shapes of 2-d LFs one needs to correct for projection.

\begin{acknowledgements}

We thank the referee for constructive comments.

\end{acknowledgements}


\begin{thebibliography}{}
\bibitem[2002]{beijersbergen_etal}
Beijersbergen, M., Hoekstra, H., van Dokkum, P. G., \& van der Hulst, J. M.,
2002, MNRAS, 329, 385 (BHDH)
\bibitem[1995]{bernstein_etal}
Bernstein, G. M., Nichol, R. C., Tyson, J. A., Ulmer, M. P., \& Wittman, D.
1995, AJ, 110, 1507 
\bibitem[1981]{fabian}
Fabian, A.~C., Hu, E.~M., Cowie, L.~L., \& Grindlay, J. 1981, ApJ,
248, 47
\bibitem[1983]{godwin}
Godwin, J. G., Metcalfe, N., \& Peach, J. V. 1983, MNRAS, 202, 113
\bibitem[1997]{lobo_etal}
Lobo, C., Biviano, A., Durret, F., Gerbal, D., Le F\`{e}vre, O.,
Mazure, A., \& Slezak, E. 1997, A\&A, 317, 385 
\bibitem[1989]{lugger}
Lugger, P. M. 1989, ApJ, 343, 572 
\bibitem[1998]{mobasher_trentham}
Mobasher, B., \& Trentham, N. D. 1998, MNRAS, 293, 315 
\bibitem[2001]{neumann_etal}
Neumann, D. M., Arnaud, M., Gastaud, R., Aghanim, N., Lumb, D., Briel,
U. G., Vestrand, W. T., Stewart, G. C., Molendi, S., \& Kittaz, J. P. D. 2001,
A\&A, 365, L74
\bibitem[1976]{schechter}
Schechter, P. 1976, ApJ, 203, 297 
\bibitem[2001]{valotto_etal}
Valotto, C. A., Moore, B., \& Lambas, D. G. 2001, ApJ, 546, 157
\bibitem[1998]{zaroubi_etal}
Zaroubi, S., Squires, G., Hoffman, Y., \& Silk, J. 1998, ApJ, 500, L87
\end{thebibliography}
\end{document}